# Operando imaging of intercalation memory in MXenes

Franz Gröbmeyer[1], Mohsen Beladi[1*], Christoph G. Gruber[1], Ruocun Wang[2,3], Pol Salles[4,5], Nhu Quynh Nguyen[3], Jakub Drnec[4], Yury Gogotsi[2*], Emiliano Cortés[1*]

[1] Nanoinstitute Munich, Faculty of Physics, Ludwig-Maximilians-Universität München, 80539 Munich, Germany.

[2] A.J. Drexel Nanomaterials Institute and Department of Materials Science & Engineering, Drexel University, 3141 Chestnut Street, Philadelphia, PA 19104, USA.

[3] Department of Materials Science & Engineering, University of North Texas, 3940 N. Elm Street, Denton 75207, USA.

[4] ESRF, The European Synchrotron, 71 Avenue des Martyrs, CS40220, 38043 Grenoble Cedex 9, France.

[5] Department of Mining, Industrial and ICT Engineering, Universitat Politècnica de Catalunya (UPC), Manresa, Spain.

* Corresponding authors: m.mousavi@physik.uni-muenchen.de, gogotsi@drexel.edu, emiliano.cortes@lmu.de

**Abstract**

Ion intercalation enables reversible control of charge, structure and function in layered solids. Yet layered materials contain nanosheets with different thicknesses and stacking, whose intercalation pathways are averaged out in ensemble measurements. This becomes critical under Å-scale confinement, where ion transport couples to solvent reorganization and host deformation, leaving the origins of kinetics, reversibility and activation unresolved. Here we combine operando interferometric scattering and interference reflection microscopy to resolve proton-driven dynamics in individual single- and multilayer $Ti_3C_2T_x$ MXene flakes and show that even a few stacked layers introduce a significant kinetic barrier. Cycling then separates nanosheets into three persistent behaviors: reversible monolayers, restacked multilayers with incomplete recovery and delayed deintercalation, and coherently stacked multilayers that self-stabilize through reversible folding. Operando synchrotron X-ray diffraction shows this memory at electrode scale, where the interlayer structure converges towards a reproducible state. Our findings reveal cycling-induced structural memory that governs subsequent intercalation in layered MXenes.

Ion intercalation – the reversible insertion of ionic species into layered or porous solids – is central to technologies ranging from electrochemical energy storage and conversion to ion-responsive sensing and actuation.[1,2] Under Å-to-nanometer confinement, ion transport becomes strongly coupled to solvent reorganization and host deformation, making kinetics and reversibility highly sensitive to local chemistry and confinement geometry.[3-5] Yet these coupled processes are largely understood through ensemble-averaged measurements, which obscure how individual nanoscale structures accommodate ion insertion.[6,7] This limitation is particularly consequential during the initial cycles, when electrochemical and structural responses can evolve substantially before becoming reproducible.[8,9] Which local processes drive this conditioning, and how they vary across individual nanoscale structures, remain unclear.

This challenge is amplified in multilayered 2D materials, where ions move through Å-scale interlayer spaces whose accessibility strongly depends on local interlayer spacing, surface chemistry, and hydration.[10-13] Materials assembled from exfoliated nanosheets are inherently heterogeneous: layer number, overlap and local stacking can vary substantially across an otherwise macroscopically uniform sample. Conventional electrochemistry averages over this heterogeneity, while operando diffraction and spectroscopy primarily report spatially averaged structural evolution.[6,7,14-16] Consequently, the nanoscale origins of the observed kinetics, reversibility, and structural conditioning remain difficult to identify directly.

Accessing these dynamics requires an operando approach that can follow individual nanosheets while remaining sensitive to nanoscale structural changes. Here, we combine interference reflection microscopy (IRM)[17] with interferometric scattering microscopy (iSCAT)[18,19] to resolve individual 2D nanosheet dynamics in real time. IRM captures nanosheet-wide changes in reflected intensity associated with intercalation-driven changes in dielectric response and thickness, while iSCAT resolves localized scattering from nanoscale structural heterogeneity superimposed on the global reflectance signal. These complementary measurements connect nanosheet-wide intercalation dynamics to localized structural changes, with sub-5-nm axial sensitivity, µs–ms temporal resolution, and <10-nm localization precision. Originally developed for biological imaging,[20-23] interferometric scattering approaches have more recently expanded into materials science and nanoscience, enabling visualization of processes ranging from framework formation and single-particle functionality to particle interactions and phase evolution. [24-34]

We apply this approach to $Ti_3C_2T_x$, the archetypal MXene and a benchmark material in the rapidly expanding field of two-dimensional carbides and nitrides.[10,35,36] Its conductive, hydrophilic layers form Å-confined galleries in which fast proton/water intercalation coexists with pronounced early-cycle activation, hysteresis and intercalation–deintercalation asymmetry.[12,37,38] Here, we show that the macroscopic signatures of MXene intercalation emerge from distinct nanosheet-level responses set by local stacking. Remarkably, only a few stacked layers are sufficient to substantially alter the intercalation kinetics, while repeated cycling drives individual nanosheets towards persistent structural configuration. Rather than responding uniformly, different stacking arrangements either show incomplete structural recovery or organize into reversible deformation modes. These nanosheet-specific modes establish a cycle- dependent mechanochemical memory that governs subsequent intercalation. Operando X-ray diffraction further shows that this history-dependent deformation extends to the electrode scale, linking local structural adaptation to the evolving interlayer state of $Ti_3C_2T_x$ films.

## Operando imaging of individual MXene nanosheets

$Ti_3C_2T_x$ MXene consists of Ti-C-Ti-C-Ti layers terminated by -O, -OH, -Cl and -F groups, which regulate interlayer hydration and interactions; the spacing expands from ~6–7 Å in the dry state to ~10–12 Å upon hydration (Fig. 1a).[39] Selective etching and delamination of $Ti_3AlC_2$ yields a heterogeneous distribution ranging from fully exfoliated single-layer to partially exfoliated, multi-layer nanosheets with different thicknesses and lateral dimensions.[40] After deposition on indium tin oxide (ITO)-coated glass, these appear as spatially separated nanosheets with locally overlapping regions (Fig. 1b and Supplementary Note 1.1; Fig. S1). Under cathodic polarization in oxygen-free 20 mM $H_2SO_4$, proton and hydronium ion insertion leads to Ti reduction and expansion of the confined interlayer structure:[41]

$$Ti_3C_2T_x + yH^+ + ye^- \rightleftharpoons H_yTi_3C_2T_x$$

Our home-built iSCAT/IRM microscope (Supplementary Note 1.2; Fig. S2) integrated with a custom three-electrode electrochemical cell (Supplementary Note 1.3; Fig. S3) detects the intensity $I$ made up from the interference of light reflected from the substrate–electrolyte interface $E_r$ and light reflected or scattered by individual $Ti_3C_2T_x$ nanosheets $E_{probe}$ (Fig. 1c). Within each nanosheet, intercalation-induced changes in dielectric response and optical thickness generate a spatially extended IRM signal, while local out-of-plane heterogeneities produce superimposed iSCAT contrast. Both contributions are recorded simultaneously, enabling nanosheet-wide and localized structural responses to be followed in operando (Supplementary Notes 1.2 and 2).

Baseline ("static") optical contrast provides a direct measure of nanosheet thickness (Fig. 1b). We modeled the glass/ITO/MXene/electrolyte stack using a multilayer Fresnel transfer-matrix method (Supplementary Note 3; Fig. S5), which predicts a non-monotonic contrast-thickness relationship arising from the transition between destructive and constructive optical interference. Calibrating against correlative atomic force microscopy (AFM) measurements (Fig. 1d, Supplementary Notes 3.2 and 3.3; Fig. S6,S7) enables layer number to be assigned directly from the optical image, as demonstrated for the monolayer, bilayer and overlapping trilayer regions in Fig. 1b.

When a potential is applied, changes in optical contrast relative to the baseline provide a dynamic measure of intercalation ("dynamic" contrast).  Figure 1e shows the response of the regions identified in Fig. 1b during a cyclic potential sweep. During the cathodic sweep, $\Delta I/I$ increases across all thicknesses and increases with increasing layer number, consistent with proton and water insertion increasing both the physical thickness and the refractive index of the interlayer galleries, thereby increasing the effective optical path length (see Supplementary Video 1, Supplementary Note 3.4 and 4.1; Fig. S8,S9). Upon reversing the potential, the contrast approaches its initial value, thereby enabling label-free, real-time tracking of intercalation in individual nanosheets. The wide-field illumination image captures several nanosheets simultaneously and enables rapid screening of many more across the same sample, providing high-throughput access to nanoscale electrochemical events in their native heterogeneous state.

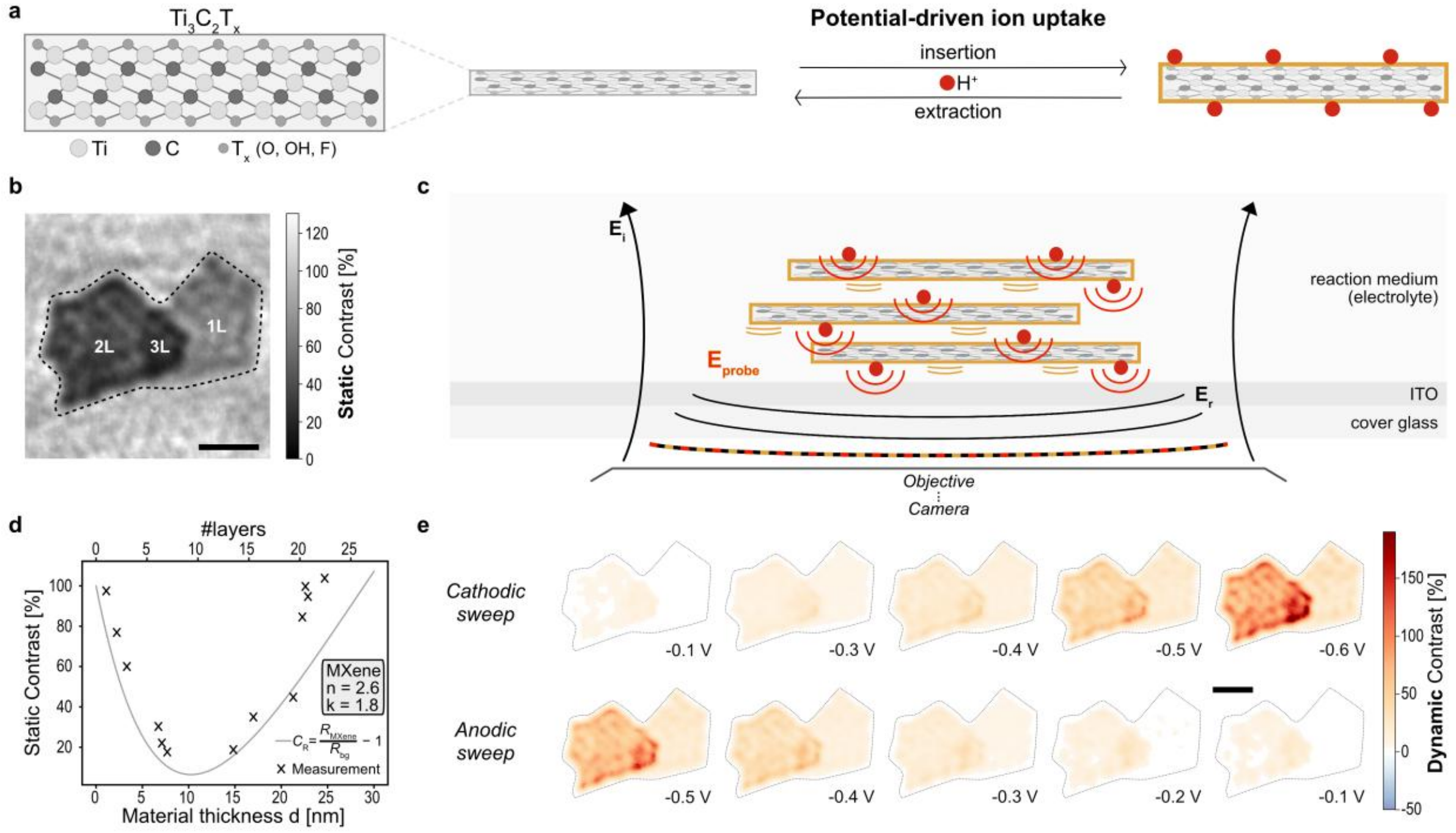

**Fig. 1 | Operando iSCAT/IRM resolves proton-driven electrochemical changes in individual MXene nanosheets. a,** Atomic structure of $Ti_3C_2T_x$, comprising Ti–C–Ti layers terminated by $T_x$ = –OH, –O and –F groups. Electrochemical polarization drives reversible proton-coupled redox at the MXene surface and, in multilayer structures, proton intercalation into hydrated interlayer galleries accompanied by changes in the confined interlayer environment. **b,** Static optical contrast image of two overlapping $Ti_3C_2T_x$ nanosheets forming regions of one, two, and three stacked layers. The contrast changes systematically with local thickness, enabling the different layer numbers to be distinguished optically; assignments were validated by correlative atomic force microscopy (AFM) measurements (Supplementary Note 3.3). **c,** Schematic of the iSCAT/IRM detection principle. Light reflected at the ITO–electrolyte interface ($E_r$) interferes with light reflected or scattered by the $Ti_3C_2T_x$ nanosheet ($E_{probe}$). Changes in dielectric response and optical thickness produce spatially extended IRM contrast, whereas localized out-of-plane heterogeneities contribute additional iSCAT contrast, enabling both nanosheet-wide and localized electrochemical changes to be followed in operando. **d,** Fresnel transfer-matrix modelled dependence of static optical contrast on $Ti_3C_2T_x$ thickness for the glass/ITO/MXene/electrolyte stack. The non-monotonic response results from the evolving interference condition with increasing layer number. Correlative AFM measurements calibrate the model and enable local layer number to be assigned from the static optical contrast. **e,** Dynamic contrast maps of the monolayer, bilayer and trilayer regions during a potential sweep from 0.1 V to −0.6 V and back to 0.1 V versus Ag/AgCl. The contrast increases during the cathodic sweep and scales with layer number, reflecting proton-coupled redox and, for multilayers, associated changes in the hydrated interlayer galleries. Reversal of the potential largely restores the initial contrast, demonstrating reversible operando tracking of the local electrochemical response. Scale bars: 2 µm.

## Layer-dependent kinetics of Å-confined proton intercalation

In macroscopic electrodes, rate dependence and hysteresis combine contributions from transport through the porous electrode and ion insertion within the active material.[42] This ambiguity is amplified by top-down delamination of etched $Ti_3C_2T_x$ multilayers, which produces a heterogeneous distribution of monolayers, few-layer fragments, and thicker aggregates that can further restack after deposition. Because charge-transfer kinetics vary with layer count,[43] ensemble measurements cannot determine which structures impose the observed kinetic limitation.

To separate these layer-dependent contributions from electrode-scale transport, we tracked $Ti_3C_2T_x$ nanosheets from 1L to 6+L across different scan rates during cyclic voltammetry (CV). The optical response, $\Delta I/I$, increases systematically from 1L to 6+L nanosheets (Fig. 2a,b; Supplementary Video 2; Supplementary Note 4.2; Fig. S10), consistent with contributions from additional proton/water-accessible galleries. Crucially, the contrast response of an individual nanosheet closely follows the integrated electrode charge during the same potential sweep (Fig. 2c), directly linking the local optical response to intercalation state. With the high acquisition rate of iSCAT/IRM, its time derivative, *dContrast/dt*, resolves these dynamics with sufficient temporal resolution to generate an optical voltammogram for every individual nanosheet (Fig. 2d; Supplementary Note 2.4), revealing the potential and rate of local insertion and deintercalation.[6,7]

Comparing 1L and 6L+ nanosheets reveals that the rate loss emerges over only a few stacked layers (Fig. 2e). Increasing the scan rate from 10 to 100 mV $s^{-1}$ suppresses the peak contrast by only ~2% for 1L but by ~36% for 6L+. Across all measured thicknesses, this suppression increases monotonically with layer count (Fig. 2f), showing that the kinetic constraint develops within the MXene stack itself, independently of porous-electrode transport.

The optical-voltammetric peaks show progressive broadening and hysteresis with increasing scan rate (Fig. 2g). We quantify this layer-dependent kinetic evolution using an optical $b$-value obtained from the scan-rate dependence of the optical-voltammetric peak, analogous to the $i_p \propto v^b$ analysis used in conventional voltammetry (Fig. 2h; Supplementary Note 2.4).[44] The response is predominantly capacitive for a monolayer ($b \approx 0.978$-cathodic; 0.972 anodic) but progressively departs from this limit with each additional layer, reaching $b \approx 0.711$ and 0.788, respectively. Thus, adding only a few Å-confined galleries is sufficient to transform a directly electrolyte-accessible, surface-like response into a finite diffusion regime in which transport and interlayer reorganization increasingly constrain the electrochemical reaction.[45]

Prior work has largely attributed rate limitations in MXenes to electrode-scale restacking and tortuous ion pathways, which can be mitigated by macroporous or vertically aligned configurations.[35,42] Our measurements reveal a distinct limitation at a much smaller length scale: A few layers of local restacking are sufficient to drive a pronounced departure from surface-controlled kinetics, despite a transport distance of only a few nanometers. This behavior cannot be explained by path length alone. Under Å-scale confinement, proton transport is coupled to a strongly modified local environment, in which water exhibits local dielectric and hydrogen-bonding properties, proton solvation is altered, and ion–host interactions depend sensitively on interlayer spacing and surface chemistry.[39,46-48] The kinetic constraint therefore seems to not arise from transport distance alone, but from the molecular reorganization and strain resulting from protonation/deprotonation of the MXene surface, imposed by successive Å-confined galleries, making layer count an intrinsic determinant of intercalation kinetics.

These layer-averaged kinetics establish confinement as an intrinsic kinetic constraint, but do not reveal how that constraint evolves locally during repeated cycling. We therefore next resolve the spatial response of individual nanosheets over time.

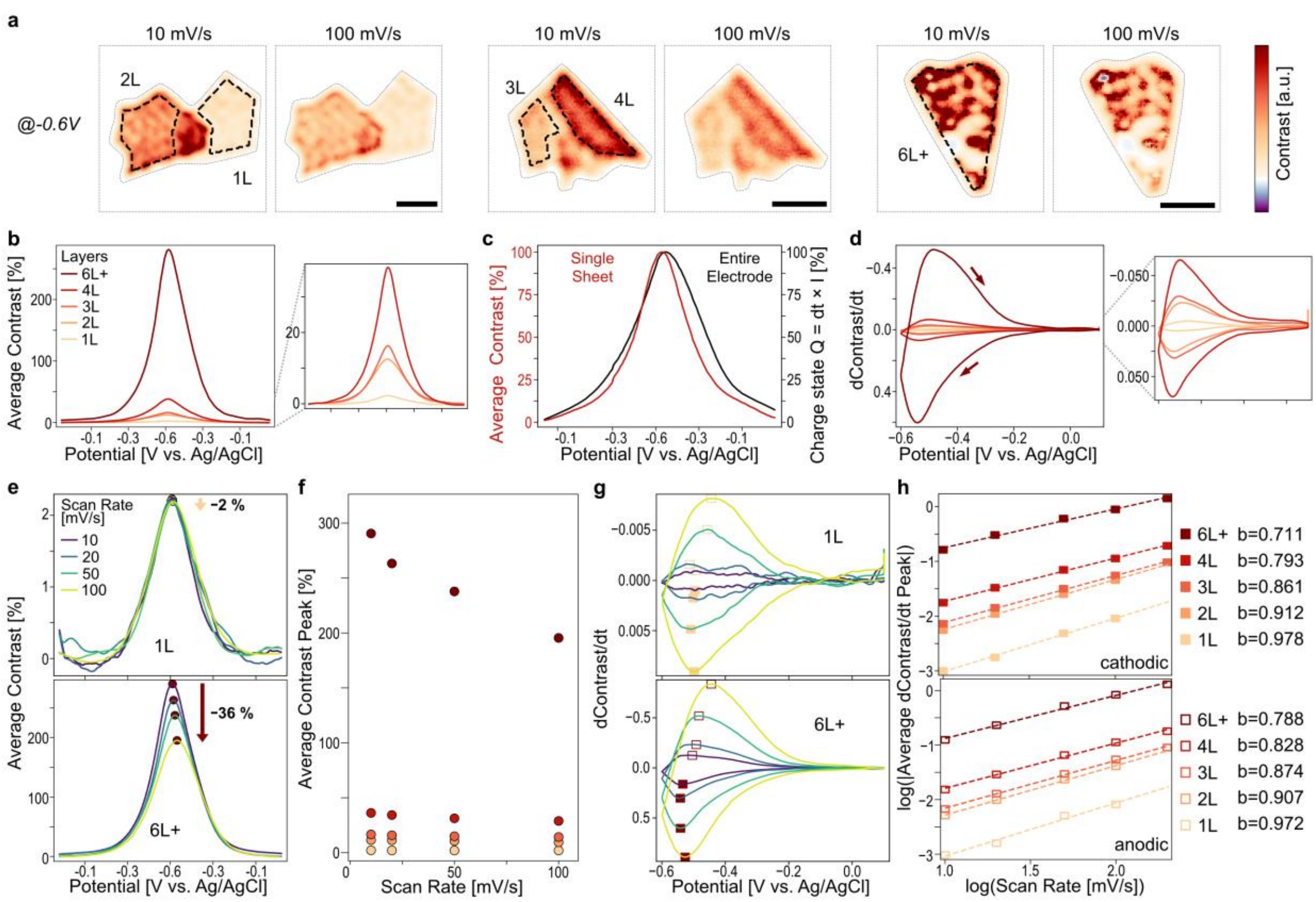

**Fig. 2 | Layer count imposes an intrinsic kinetic constraint on proton intercalation. a,** Dynamic contrast maps of representative $Ti_3C_2T_x$ nanosheets containing 1L/2L, 3L/4L and 6L+ regions, recorded at the most intercalated state (−0.6 V versus Ag/AgCl) at scan rates of 10 and 100 mV s⁻¹. Optical contrast increases with layer number, whereas the response of thicker regions is increasingly suppressed at the higher scan rate. **b,** Nanosheet-averaged optical contrast as a function of potential for different layer counts at 50 mV s⁻¹, showing a nonlinear increase in peak contrast with increasing thickness. Inset, enlarged response of 1L–4L nanosheets. **c,** Comparison during a single potential sweep of the average optical contrast of an individual nanosheet and the integrated charge of the macroscopic electrode. Their close correspondence establishes the local optical contrast as a proxy for the intercalation state of an individual nanosheet. **d,** Time derivative of the optical contrast, *dContrast/dt*, for the same layer counts as in **b**, providing an optical voltammogram that resolves the potential-dependent rate of the local electrochemical response. Inset, enlarged response of 1L–4L nanosheets. **e,** Scan-rate dependence of the contrast–potential response for 1L (top) and 6L+ (bottom) nanosheets. Increasing the scan rate from 10 to 100 mV s⁻¹ suppresses the peak contrast by only ~2% for 1L but by ~36% for 6L+. **f,** Peak optical contrast as a function of scan rate for all layer counts, showing progressively stronger rate-dependent suppression with increasing layer number. **g,** Corresponding optical voltammograms for 1L (top) and 6L+ (bottom). The monolayer response remains comparatively narrow across scan rates,

whereas the 6L+ response broadens and shifts, revealing increasingly constrained kinetics within the multilayer stack. **h,** Log–log dependence of the cathodic (top) and anodic (bottom) optical-voltammetric peak magnitude, |*dContrast/dt*|, on scan rate. Linear fits yield the optical kinetic *b*-value, which decreases from 0.978 and 0.972 for 1L to 0.711 and 0.788 for 6L+ on the cathodic and anodic sweeps, respectively, demonstrating a progressive departure from predominantly surface-controlled kinetics with increasing layer number. Scale bars: 2 μm.

## Evolution of intercalation memory

Early-cycle activation is widely reported in MXene electrodes, including $Ti_3C_2T_x$, and is commonly associated with progressive gallery opening, hydration and improved ion accessibility (Supplementary Note 5; Fig. S11,S12).[46,49] Yet activation is observed at the ensemble level and does not reveal the microscopic changes that produce it. This is particularly important because cycling is thought to improve access to the same interlayer galleries that impose the kinetic limitation identified above.

We therefore examine a dilute MXene population in which monolayers, few-layer stacks and overlapping regions remain optically distinguishable within a common electrochemical environment (Fig. 3a). Dynamic contrast at the most intercalated state reveals pronounced spatial heterogeneity across the population (Fig. 3b), while the electrode-level response evolves markedly over the first 200 cycles (Fig. 3c). What, then, is actually changing within the layered structure during activation? To answer this, we track individual nanosheets across repeated potential sweeps and identify recurring behaviors.

**Monolayers**. Monolayer $Ti_3C_2T_x$ shows essentially unchanged static contrast and morphology over hundreds of cycles, indicating no persistent structural reorganization (Fig. 3d). The dynamic contrast maps show that, in both the first (#6) and final (#535) cycles, the contrast increases uniformly across the nanosheet at −0.6 V and returns nearly to its initial level at −0.2 V on reversal, demonstrating a spatially homogeneous and highly reversible proton-coupled redox response (Supplementary Video 3). Consistently, the contrast–potential curves retain nearly identical shape and position over several hundred cycles. A slight decrease in overall contrast during the first few cycles is attributed to the equilibration of processing-derived surface species, a behavior commonly seen in early cycling of delaminated MXenes.[37] Without confinement between MXene layers, the monolayer therefore provides a highly reproducible and reversible reference against which the cycling response of multilayer nanosheets can be assessed.

**Restacked and overlapping multilayers**. Restacked and overlapping configurations are common in processed MXene films, where incomplete delamination and subsequent deposition generate regions with different local thickness and stacking configuration. Unlike the monolayer, the static contrast reveals a persistent change with cycling: the contrast profile changes substantially over 300 cycles, indicating that parts of the multilayer do not fully return to their initial state after repeated sweeps (Fig. 3e). The dynamic maps reveal how this change develops. At cycle 4, regions II–IV show a pronounced contrast increase at −0.6 V that largely disappears again by −0.2 V on the reverse sweep, indicating a large and mostly reversible response (Supplementary Video 4). By cycle 332, the response at −0.6 V is markedly weaker, whereas substantially more contrast remains at −0.2 V, indicating increasingly incomplete recovery. This evolution is strongly local: region I changes little, while neighboring regions II–IV evolve to varying degrees. The region-resolved contrast curves make this redistribution particularly clear: at cycle 4 the response is dominated by a sharp peak around −0.6 V, reaching almost 500% in region IV, whereas after more than 300 cycles this peak is strongly suppressed across the affected regions and a larger fraction of the response remains away from the peak potential.

The contrast–potential response over successive cycles further show that this is a gradual process: the main intercalation peak progressively decreases while shoulders develop on the cathodic and, particularly, anodic branches, revealing an increasingly asymmetric ion insertion-extraction behavior. We interpret these combined signatures as incomplete relaxation of the confined interlayer state. Proton insertion and the associated reorganization of confined water modify the galleries during the cathodic sweep, but on the reversal, part of this confined state appears not to relax or remove all intercalants on the same timescale.

Repeated cycling therefore leaves an increasingly persistent interlayer configuration, reducing the reversible structural response available in subsequent cycles and producing delayed deintercalation. Thus, rather than simply opening during activation, common restacked MXene configurations can progressively retain part of their intercalated state and lose reversibility with cycling.

**Coherently stacked multilayers**. A third response is observed in partially exfoliated, stand-alone multilayer nanosheets with uniform thickness, where several MXene sheets remain coherently stacked but are separated from surrounding nanosheets rather than forming a restacked network. In these nanosheets, the static contrast remains essentially unchanged over hundreds of cycles, in marked contrast to the restacked multilayers (Fig. 3f). The dynamic maps, however, reveal a striking reorganization of the intercalation response (Supplementary Video 5). At cycle 5, the contrast at −0.6 V is comparatively uniform and largely disappears again at −0.2 V on reversal. By cycle 534, the response at −0.6 V has reorganized into narrow, highly localized features, which again vanish when the potential returns to −0.2 V. Thus, cycling creates a new spatial pattern of reversible, fold-like out-of-plane deformation that appears only in the intercalated state.

Remarkably, the contrast–potential curves remain smooth and highly reproducible throughout this evolution, showing that a nearly spatially averaged optical response can conceal a profound reorganization of where and how the multilayer accommodates intercalation. We therefore assign these features to reversible fold formation, in which coherently stacked layers progressively concentrate intercalation-induced strain into defined out-of-plane deformation channels rather than retaining a permanently expanded structure.

Cycling can therefore reorganize a multilayer into a mechanically encoded, reversible intercalation state without substantially altering its average electrochemical response.

The key implication is that activation is not a uniform relaxation towards a more accessible state, but a history-dependent reorganization of the confined multilayer structure. Repeated cycling can therefore change the future response of a nanosheet even when its average electrochemical signature appears stable, thereby establishing electrochemical history as a state variable governing subsequent intercalation. Reversible fold formation provides a direct physical manifestation of this coupling between ion insertion and mechanical reorganization, shown by localized out-of-plane deformations, which we examine next in more detail.

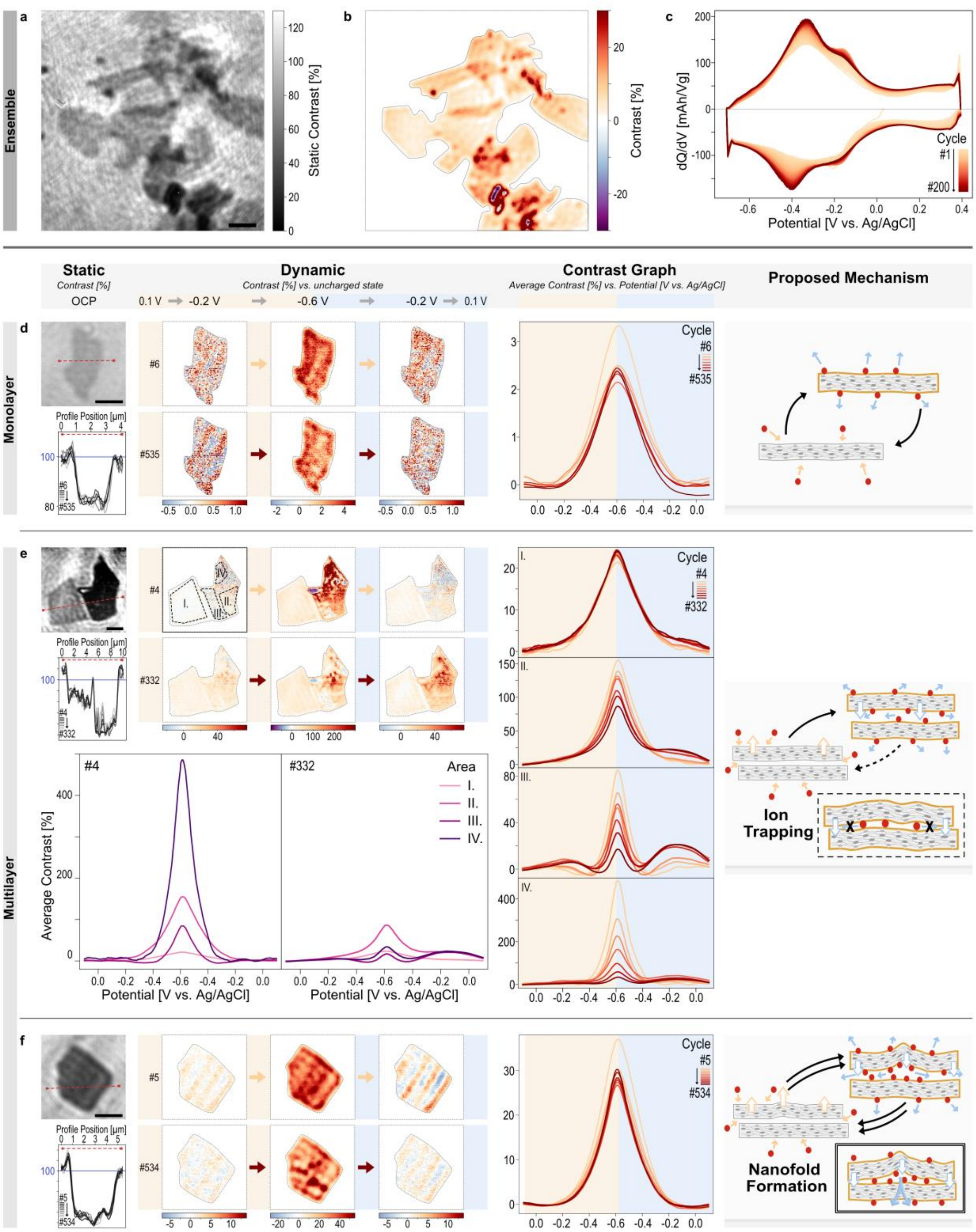

**Fig. 3 | Cycling imprints stacking configuration-dependent intercalation memory in individual MXene nanosheets. a, b,** Static contrast image of a sparsely covered region containing coexisting monolayers, multilayers and partially overlapping nanosheets within a single field of view, reproducing the heterogeneous stacking configurations of a macroscopic drop-cast film at reduced density. **a,** Static contrast image showing the heterogeneous local stacking configurations. **b,** Dynamic contrast map of the same region at −0.6 V versus Ag/AgCl, corresponding to the most intercalated state, showing pronounced spatial heterogeneity in the electrochemical response. **c,** Electrode-level differential charge response, *dQ/dV*, recorded over 200

consecutive cycles, showing the progressive evolution of the cathodic and anodic electrochemical features during cycling. **d–f,** Representative recurrent responses resolved by tracking individual nanosheets over hundreds of potential sweeps. For each example, the static contrast image and line profile show the morphology and its evolution with cycling; dynamic contrast maps show the response at −0.6 V and after reversal to −0.2 V for an early and a late cycle; contrast–potential curves show the corresponding evolution over successive cycles; and schematics illustrate the proposed microscopic response. **d,** Monolayer $Ti_3C_2T_x$. The dynamic response remains spatially uniform and returns close to its initial state on reversal, while the contrast–potential curves remain highly reproducible over more than 500 cycles, consistent with reversible proton-coupled redox without confinement between MXene layers. **e,** Restacked and overlapping multilayers. Four regions (I–IV) within the same assembly exhibit distinct cycling responses. Region I remains comparatively stable, whereas regions II–IV show a progressive decrease in the response at −0.6 V, increasing residual contrast at −0.2 V and the emergence of cathodic and anodic shoulders with cycling. The static contrast profile also evolves between the early and late cycles, consistent with incomplete relaxation and progressive retention of the confined interlayer state. **f,** Coherently stacked multilayer. An initially comparatively uniform response reorganizes during cycling into localized fold-like features at −0.6 V that reappear again at −0.2 V. The waves in the early cycle are due to fluctuations nearby, presumably from loosely attached material. Despite this pronounced spatial reorganization, the spatially averaged contrast–potential response remains highly reproducible, consistent with reversible accommodation of intercalation-induced strain through fold formation. Scale bars: 2 µm.

## Reversible fold dynamics

The reversible folds identified above reveal that intercalation in stacked MXenes can be accommodated through spatially localized mechanical out-of-plane deformation rather than uniform interlayer expansion and contraction. Because monolayers remain morphologically stable, this deformation points to a collective response of coupled MXene sheets and their confined interlayer galleries. Operando iSCAT/IRM resolves how this mechanical response is progressively reorganized by cycling, revealing two distinct regimes: macrofolds and nanofolds (Fig. 4).

**Macrofolds**. In some stacked multilayers, local variations in stacking and residual strain create mechanically susceptible regions where intercalation-induced stress concentrates, leading to large out-of-plane deformations. These macrofolds can therefore be present upon exposure to electrolyte or emerge during the first cycles. In the example shown, the static images reveal a large high-contrast fold that persists after the initial sweep but progressively contracts and reorganizes with cycling (Fig. 4a). The corresponding dynamic contrast line profiles distinguish persistent structural reorganization from the deformation induced within each potential sweep (Fig. 4b). For each sweep, the lower and upper yellow traces correspond to 0.1 V before and after cycling, whereas the purple trace shows the intercalated state at −0.6 V. In the early sweeps, the two 0.1 V profiles differ markedly, revealing incomplete structural recovery; with cycling they progressively converge. Simultaneously, the −0.6 V profile changes substantially, indicating that conditioning not only removes persistent deformation but also redistributes the locations where the stack deforms during intercalation (Fig. 4b; Supplementary Video 6).

The dynamic maps resolve this conditioning directly. During the early sweep, intercalation produces strongly localized deformation between −0.4 and −0.6 V, but the reversal does not fully restore the starting configuration, leaving a pronounced, spatially heterogeneous residual contrast at 0 V (Fig. 4c,d). By sweep iii, the intercalation response is more reproducibly distributed across the nanosheet and largely disappears on reversal, indicating substantially improved mechanical recovery (Fig. 4e,f). Macrofold conditioning therefore converts an initially persistent, poorly recovering deformation into a reproducible potential-driven response: repeated intercalation progressively selects preferred deformation modes that can subsequently open and relax reversibly.

**Nanofolds.** A second fold regime emerges in coherently stacked multilayers that remain morphologically flat between cycles but progressively develop a spatial pattern of deformation only during intercalation. In the representative 1L–2L–3L nanosheet, the static images remain essentially unchanged throughout conditioning, with no persistent folds visible after each sweep (Fig. 4g). The line profiles through the 2L region separate this stable baseline from the evolving intercalated state (Fig. 4h). For each sweep, the profiles at −0.3 V before and after intercalation remain nearly identical, whereas the profile at −0.6 V progressively evolves from a broad change in contrast into several spatially resolved peaks. Cycling therefore reorganizes the deformation occurring within the intercalated state without leaving a permanent structural imprint between sweeps (Supplementary Video 7).

The dynamic maps directly visualize this transition. During the early sweep, the contrast evolves relatively uniformly across the nanosheet as the potential approaches −0.6 V and disappears again on reversal (Fig. 4i,j). The faint, line-like feature visible across several potentials in this early sweep is attributed to the underlying ITO substrate, as observed through the ultrathin MXene, rather than to fold formation. After extended cycling, the same potential excursion produces a series of narrow stripe-like features that intensify towards −0.6 V and vanish again on returning to −0.3 V (Fig. 4k,l). Their lateral positions remain largely preserved across subsequent sweeps, indicating that conditioning establishes stable, potential-dependent nanofolds that repeatedly accommodate intercalation-induced deformation and fully relax upon reversal. Notably, no comparable nanofold pattern develops in the adjacent 1L region, providing direct evidence that these folds require multiple stacked layers and arise from mechanical coupling between stacked MXene sheets. Unlike macrofolds, which involve an initially persistent deformation that is progressively conditioned, nanofolds therefore emerge from an apparently unchanged multilayer and exist only within the intercalated state.

These two apparently different responses reveal the same underlying adaptation: cycling progressively localizes intercalation-induced strain into preferred out-of-plane deformation modes. Proton insertion reorganizes the confined proton–water environment and changes the interlayer spacing, while ion intercalation is known to strongly modify the out-of-plane mechanics of $Ti_3C_2T_x$.[37,46,50,51] Because mechanically coupled MXene sheets cannot accommodate these local dimensional changes independently, we propose that mechanical mismatch develops and is relieved by buckling at favorable sites. Repeated cycling then selects and stabilizes the deformation modes that most reversibly accommodate this mismatch.

Crucially, partially delaminated, coherently stacked multilayers can use this mechanism to self-stabilize. Instead of accumulating incomplete structural recovery and delayed proton deintercalation, as in the restacked multilayers of Fig. 3e, they channel strain into reversible folds, recovering their initial configuration after each sweep. Folding is therefore an adaptive strain-relief mechanism that preserves reversible intercalation. When stacking is disordered or mechanically constrained, this adaptation is frustrated, favoring persistent deformation and delayed relaxation.

Local stacking determines whether intercalation-induced deformation relaxes reversibly or persists after cycling. If this cycle-selected response is repeated across the electrode, it should emerge as an ensemble-scale structural memory, which we examine next.

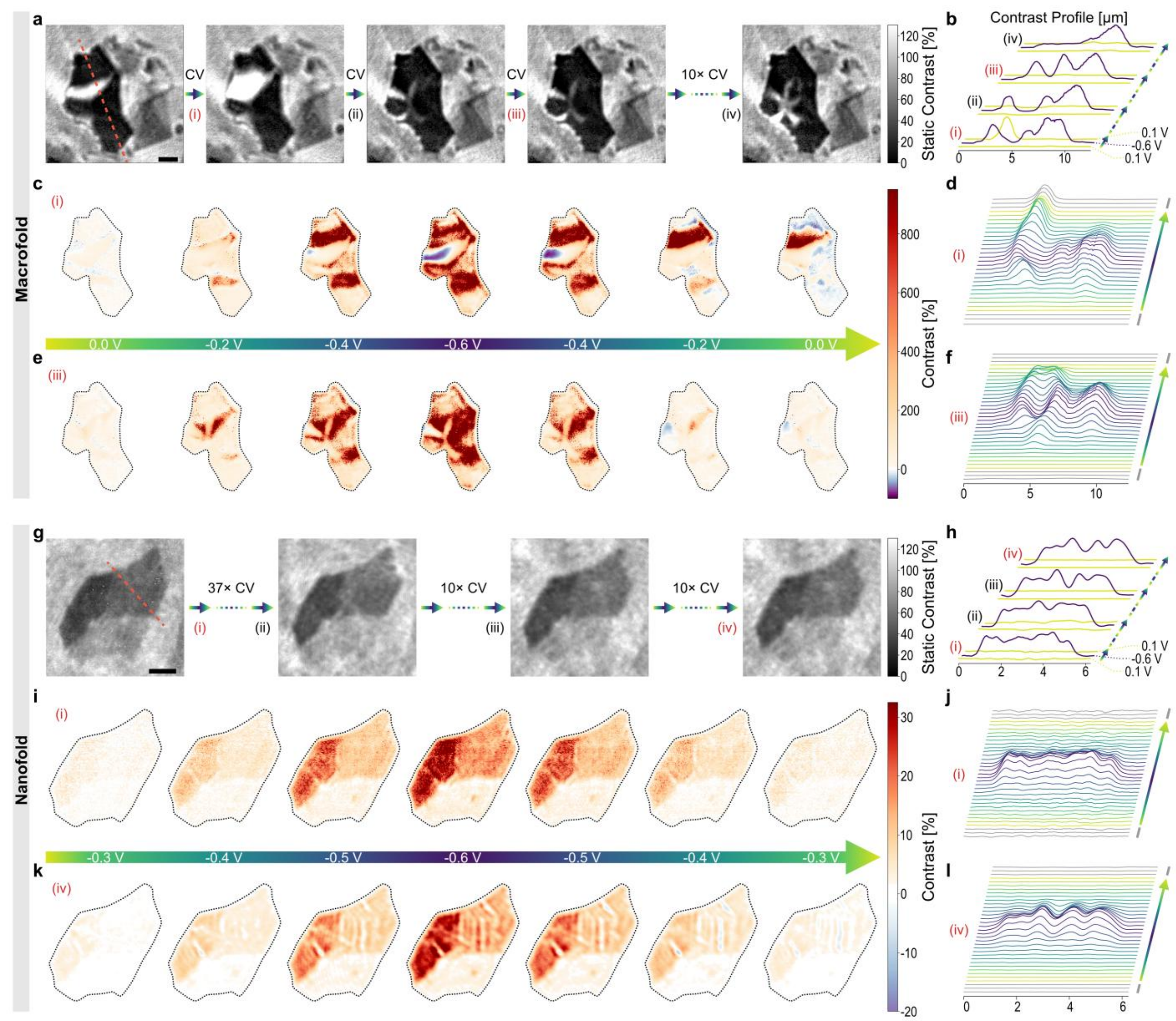

**Fig. 4 | Cycling reorganizes intercalation-induced deformation into reversible macro- and nanofolds. a–f, Macrofold evolution. a**, Static contrast images of a representative multilayer containing a pronounced macrofold at successive stages of electrochemical conditioning (*i–iv*). The initially extended deformation progressively contracts and reorganizes with cycling. **b**, iSCAT/IRM contrast profiles along the dashed line in a for states *i–iv* (Potential [V vs. Ag/AgCl]). The lower and upper yellow traces show 0.1 V before and after each sweep, respectively; blue traces show the maximally intercalated state at −0.6 V. The initially distinct yellow profiles progressively converge with cycling, indicating improved structural recovery, while the evolving blue profile shows redistribution of the intercalation-induced deformation into a more reproducible configuration. **c**, Potential-resolved dynamic contrast maps during an early sweep (*i*), showing strongly localized deformation and incomplete recovery on reversal. **d**, Corresponding dynamic contrast line profiles, revealing the strongly history-dependent response. **e**, Potential-resolved dynamic contrast maps during a conditioned sweep (*iii*), showing spatially

reorganized deformation that largely reverses on return. **f**, Corresponding line profiles, demonstrating the transition to reproducible potential-driven fold dynamics. **g–l, Nanofold evolution.** **g**, Static iSCAT/IRM contrast images of a representative nanosheet containing adjacent 1L, 2L and 3L regions during successive stages of conditioning (*i*–*iv*). The baseline morphology remains essentially unchanged despite the emergence of a strongly patterned response during intercalation. **h**, Contrast profiles through the 2L region along the dashed line in g for states *i*–*iv*. For each state, the lower and upper yellow profiles correspond to −0.3 V before and after the potential sweep, respectively, whereas the blue profile corresponds to the maximally intercalated state at −0.6 V. The yellow profiles remain nearly coincident throughout conditioning, demonstrating reversible recovery, while the blue profile progressively evolves from a broad change in contrast into multiple spatially resolved peaks, revealing the emergence of nanofolds within the intercalated state. **i**, Potential-resolved dynamic contrast maps during an early sweep (*i*), showing a comparatively uniform response across the multilayer regions. **j**, Corresponding dynamic contrast line profiles, revealing broad intercalation-induced deformation without distinct folds. **k**, Potential-resolved dynamic contrast maps during a conditioned sweep (*iv*), showing narrow stripe-like features that emerge towards −0.6 V and disappear again on reversal; no comparable pattern develops in the adjacent monolayer region. **l**, Corresponding line profiles, showing spatially localized, reversible nanofolds at reproducible positions. Scale bars: 2 μm.

## Ensemble structural memory

To determine whether the conditioning and structural memory resolved in individual nanosheets persist at the electrode scale, we performed operando synchrotron grazing-incidence X-ray diffraction (GIXRD) during the initial cycling of a ~3 μm $Ti_3C_2T_x$ film during its first tens of electrochemical cycles (Supplementary Note 6; Fig. S13). This provides an ensemble view of how the stacked structure reorganizes during conditioning and whether it converges towards a reproducible structural configuration.

During each potential sweep, we tracked the $Ti_3C_2T_x$ *002* reflection, whose position reports the average out-of-plane spacing of the stacked galleries ($Q = 2\pi/d$; lower Q corresponds to larger spacing). The cycle-resolved Q maps reveal a pronounced evolution during conditioning (Fig. 5a–c). In the first cycle, the *002* peak undergoes a large shift to lower Q during the cathodic sweep, consistent with substantial interlayer expansion, but only partially returns on reversal. By cycles 11 and 25, both the Q excursion and the ion insertion-extraction hysteresis are markedly reduced, and the structural response becomes increasingly reproducible around a new baseline state. Thus, early cycling not only changes the mean structure of the film, but also reduces the magnitude of its subsequent structural breathing.

This evolution is quantified by the fitted *002* peak position (Fig. 5d). During the first cycles, the film follows a strongly hysteretic spacing evolution: the galleries expand on the cathodic sweep but do not recover their initial contracted state on reversal. With conditioning, the loop progressively narrows, and the starting and ending positions converge, while the overall Q excursion decreases. The film therefore does not return to its pristine stacking configuration; instead, early cycling establishes a new, less strongly breathing structural configuration from which subsequent intercalation becomes increasingly reproducible.

The radial width of the *002* reflection provides a complementary measure of the distribution of interlayer configurations (Fig. 5e). During conditioning, the full width at half maximum along Q (Q-FWHM) of the deintercalated state progressively narrows, indicating that the film converges towards a more uniform contracted configuration. In contrast, the intercalated state remains broader and becomes slightly more heterogeneous with cycling. Thus, conditioning establishes a reproducible structural asymmetry in which deintercalation returns the stack to a relatively well-defined state, whereas intercalation accesses a broader distribution of gallery configurations.

The azimuthal width $\chi$-FWHM captures the evolution of the film texture (Fig. 5f). It increases most strongly during the earliest cycles and then approaches a plateau, indicating that the dominant increase in the orientational spread of stacked domains occurs during conditioning. The film therefore not only changes its average interlayer spacing, but also rewrites its out-of-plane texture before settling into a more stable configuration, consistent with the cycling-induced mechanical reorganization resolved by iSCAT/IRM at the single-nanosheet level.

The key implication is that conditioning changes the structural configuration from which subsequent intercalation proceeds. The ensemble response therefore does not reflect repeated expansion and contraction of a fixed stack, but a history-dependent structure that evolves during the first cycles before becoming reproducible. This complements the iSCAT/IRM observations: the local reorganization of stacked nanosheets is retained at the film level as a cycle-dependent structural configuration. Thus, intercalation memory extends from individual confined nanosheets to the electrode scale.

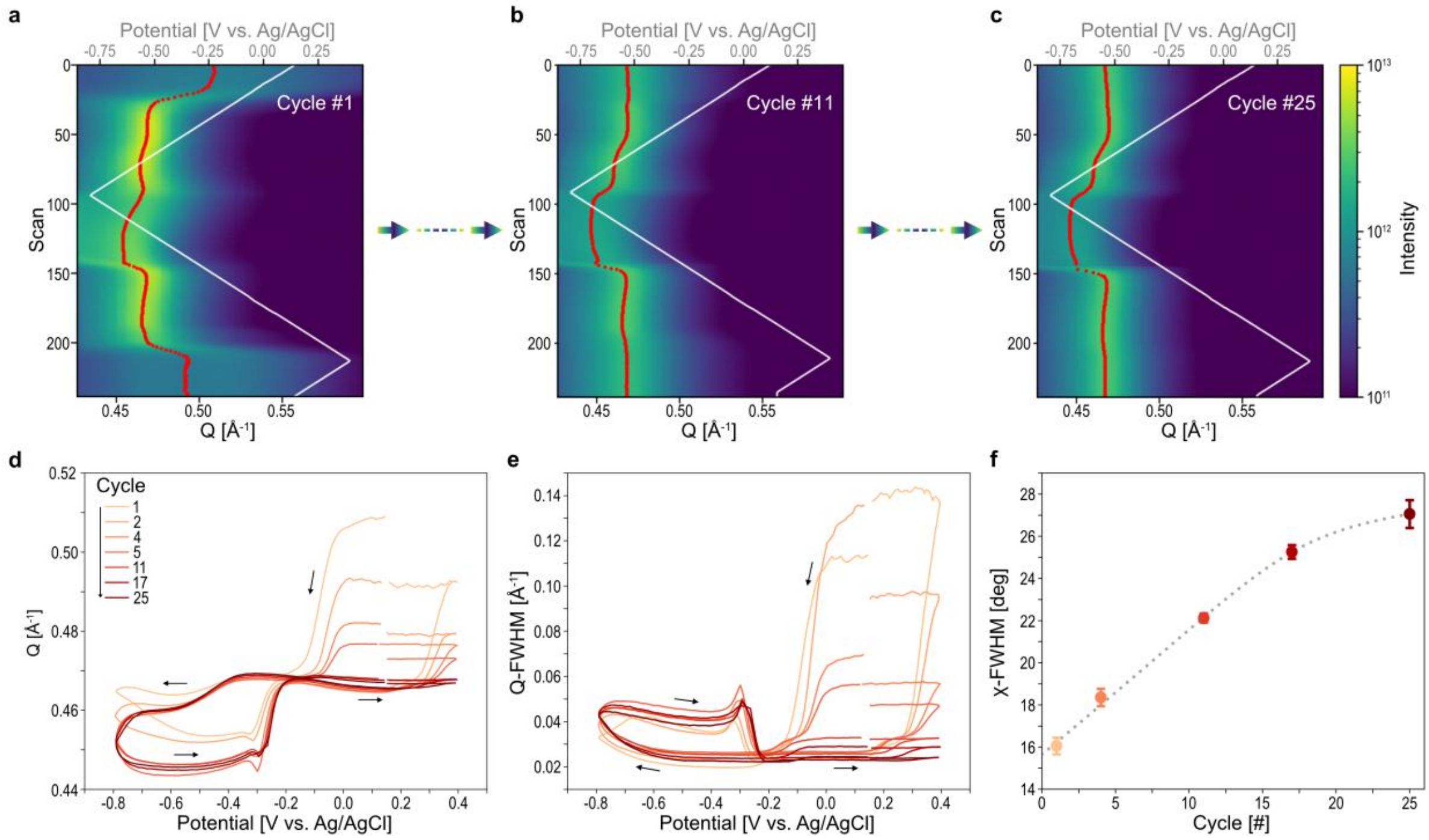

**Fig. 5 | Operando GIXRD reveals ensemble-scale structural memory during conditioning. a–c,** Operando Q–scan maps of the $Ti_3C_2T_x$ *002* reflection during cycles 1, 11 and 25, respectively. Color denotes diffraction intensity, the white line shows the applied potential, and the red line traces the fitted *002* peak position. The large, hysteretic Q excursion in the first cycle progressively decreases with conditioning, yielding a smaller and more reproducible structural response by cycle 25. **d,** Fitted *002* peak position as a function of potential for selected cycles. The pronounced spacing hysteresis and incomplete return towards the initially contracted state diminish progressively as the film converges towards a new structural response. **e,** Corresponding radial peak width (Q-FWHM), reporting changes in the distribution of out-of-plane spacings, microstrain and stacking coherence. Conditioning produces a strongly state-dependent but increasingly reproducible peak-width response. **f,** Azimuthal peak width (χ-FWHM) as a function of cycle number, showing a pronounced increase during the initial cycles

followed by a slower evolution, consistent with persistent deformation of the film texture. Arrows in d and e indicate the potential-sweep direction.

## Conclusions

In summary, we have demonstrated that operando optical imaging can resolve intercalation dynamics and structural adaptation within individual $Ti_3C_2T_x$ nanosheets. The emergence of persistent, stacking-dependent responses pathways shows that local stacking governs how confined structures respond to repeated ion insertion. Furthermore, operando GIXRD reveals that this cycle-dependent structural memory extends from individual nanosheets to the electrode scale. By linking local configuration to ion-transport kinetics, deformation and memory, our approach provides a framework for rational materials synthesis to control how layered and nanoconfined structures adapt to repeated ion intercalation.

## Methods for single-nanosheet imaging of proton intercalation

**MXene synthesis and sample preparation**. $Ti_3C_2T_x$ MXene nanosheets were synthesized using a previously described procedure.[40,52] Briefly, we selectively etched an in-house-made $Ti_3AlC_2$ MAX phase using a mixture of HCl and HF (1 g MAX phase in 20 mL of a 6:3:1 volume ratio of 12 M HCl [Fisher Scientific], DI water, and 50 wt% HF [Acros Organics]) at 35 °C for 24 hrs. The resulting multilayer powder was delaminated by mechanical shaking for 16 mins after mixing with 0.5 M LiCl (50 mL for 1 g of MAX). After washing away the excess LiCl in the solution by centrifugation at 2550g for 10 min, the single-layer and few-layer MXenes were collected by centrifugation at 7230g for 3 min. The procedure yields a stable colloidal suspension of $Ti_3C_2T_x$ MXene that is predominantly single-layer with a small fraction of few-layer nanosheets.

For optical and electrochemical measurements, delaminated nanosheets were deposited by drop-casting onto high-precision glass coverslips (thickness No. #1.5H, ~170 µm thick, Thorlabs) coated with a 100 nm indium tin oxide (ITO) film sputtered in-house. After deposition, samples were dried in a vacuum desiccator to promote adhesion between the nanosheets and the ITO surface. To meet the requirements of both single-nanosheet optical isolation and sufficient active-material coverage for ensemble electrochemical readout, regions of different nanosheet density were prepared on the same substrate: sparse areas containing optically isolated monolayer and few-layer nanosheets served as the imaging field of view, while adjacent higher-coverage regions provided the electrochemically active area for current measurements (Supplementary Fig. S1).

For operando synchrotron measurements, a freestanding $Ti_3C_2T_x$ MXene film was prepared by vacuum filtration of the colloidal suspension. This process yields a ~3 µm thick film composed of stacked MXene nanosheets with c-axis oriented out-of-plane.

**Operando interferometric scattering / interference reflection microscopy iSCAT/IRM**. Operando optical measurements were performed on a home-built inverted wide-field reflection microscope configured for interferometric scattering (iSCAT). A continuous-wave diode laser ($\lambda$ = 785 nm, typical irradiance ~10 W $cm^{-2}$ unless otherwise noted, Supplementary Note 7.1) was collimated into the back focal plane of a high-numerical-aperture oil-immersion objective (100×, NA 1.4), providing wide-field illumination over an approximately 12 × 12 $\mu m^2$ field of view. The same hardware simultaneously supports interference reflection microscopy (IRM): the distinction between the two imaging modes lies not in the optical layout, but in which part of the detected signal is interpreted. Flat, laterally uniform nanosheet terraces (lateral dimensions $\gg \lambda$) return predominantly specular reflection and are well described by the coherent reflectance of the glass/ITO/MXene/electrolyte multilayer (IRM regime). When sub-wavelength structural heterogeneities break the lateral uniformity - for example, folds, wrinkles or edge terminations - they contribute an additional scattered field that interferes with the reflected reference field, generating the characteristic iSCAT signal as a localized perturbation on the global reflectance background (Supplementary Note 1.2 and Note 2).

The combined reflected and scattered fields were collected through the same objective and imaged onto a high-speed CMOS camera (pixel size ~31 nm). Frame rates were set between 200 and 1000 fps depending on the electrochemical scan rate, with exposure times adjusted to avoid detector saturation. This configuration enables detection of refractive-index and effective-thickness variations on the order of $\Delta I/I < 1\%$, providing quantitative tracking of nanoscale ion-induced changes in $Ti_3C_2T_x$ during operando cycling. Reproducibility across nanosheets of varying thickness and stacking was additionally verified by comparing contrast–potential responses across multiple samples (Supplementary Note 7.2, Fig. S14).

**Electrochemical cell and measurement protocol**. A custom three-electrode cell was designed to be compatible with high-NA inverted microscopy (Supplementary Fig. S2). The working electrode (WE) was the ITO-coated coverslip carrying the MXene deposit, with an Ag/AgCl (3 M KCl) reference electrode (RE) and a Pt wire counter electrode (CE) completing the cell. The electrolyte for all operando optical measurements was 20 mM $H_2SO_4$, deaerated by nitrogen bubbling for at least 15 min prior to each experiment to suppress parasitic oxygen reduction.

To ensure reliable imaging, special care was taken to prevent nanosheet drift or detachment during electrolyte introduction: dried samples were kept under reduced pressure until use, and the cell was filled slowly to avoid shear forces on the deposit. Potential sweeps (cyclic voltammetry) were performed using a Metrohm µStat400 potentiostat at scan rates of 5-200 mV/s. All potentials are reported versus Ag/AgCl (3 M KCl). Electrochemical data acquisition was synchronized with camera frame capture via custom software to enable frame-by-frame assignment of applied potential.

Two complementary contrast definitions were employed (Supplementary Note 2.1). **Static contrast**, $C_s = I_{probe}/I_{bare} - 1$, references the nanosheet intensity to a nearby bare substrate region and provides a quantitative thickness calibration **Dynamic contrast**, $C(t) = I(t)/I_0 - 1$, references each frame to the pre-sweep baseline image and tracks operando intercalation-induced changes in the effective dielectric response, complex refractive index and interlayer spacing.

**Image processing and contrast analysis** (Supplementary Note 2.2). Recorded raw image stacks were processed using custom Python scripts. The pipeline comprised: (i) spatial binning (2 × 2 pixels) and temporal averaging of two consecutive frames to reduce shot noise; (ii) a rolling-shutter correction to suppress horizontal stripe artifacts inherent to the CMOS readout at high frame rates; and (iii) frame-by-frame normalization to a bare-ITO background region within the same field of view to correct for low-frequency laser-intensity fluctuations.

For each dataset, regions of interest (ROIs) were selected on interior terrace areas away from nanosheet edges to suppress contributions from boundary motion and edge-scattering artifacts (Supplementary the S2.3). The ROI-averaged contrast was smoothed with a Savitzky–Golay filter whose window size was matched to the camera sampling interval and the characteristic timescale of the potential sweep. The time derivative *dContrast/dt* was computed (equivalently, Savitzky–Golay derivative coefficients) and mapped onto the applied potential using the constant scan rate $v$, yielding an 'optical voltammogram' (*dContrast/dt* vs. E) that highlights the potentials of fastest optical change (Supplementary Note 2.4). To compare kinetics across scan rates and thicknesses, a power-law scaling analysis was applied to the derivative extrema: $(dContrast/dt)_{\mathrm{peak}} \propto v^b$, where $b$ is obtained from the slope of $\log((dContrast/dt)_{\mathrm{peak}})$ versus $\log(v)$ and distinguishes surface-limited ($b \approx 1$) from diffusion-limited scaling. For spatially resolved kinetics, pixel-wise derivative maps were computed to identify local hotspots of accelerated optical change within individual nanosheets.

**Thickness calibration**. Nanosheet thickness was determined by correlating the static optical contrast with atomic force microscopy (AFM) height profiles acquired on the same nanosheets under ambient conditions (Supplementary Note 3.3). This empirical calibration was cross-validated against the transfer-matrix (Fresnel) model of the glass/ITO/MXene/medium stack using literature optical constants for $Ti_3C_2T_x$ (Supplementary Note 3), which predicts a non-monotonic contrast–thickness relation arising from the interplay of destructive and constructive interference in the thin-film stack. Mono- to few-layer MXene nanosheets can be assigned a thickness from their baseline (static) contrast before electrochemical cycling.

**Optical modelling** (Supplementary Note 3). The reflectance of the layered glass/ITO/$Ti_3C_2T_x$/electrolyte system was computed using a transfer-matrix formalism for coherent multilayers, following Byrnes' approach and employing published optical constants for $Ti_3C_2T_x$ at 785 nm. The model calculates the complex reflection coefficient r for the stack and expresses the contrast as $C = R/R_0 - 1$, where $R_0$ is the bare-substrate reflectance. Perturbation sweeps over changes in the real refractive index ($\Delta n$), extinction coefficient ($\Delta k$) and interlayer spacing ($\Delta d$) were used to disentangle the optical contributions of dielectric, absorptive and geometric changes appearing during intercalation. This model describes only the laterally uniform (IRM) response; sub-diffraction-limit features such as nanofolds and wrinkles break the thin-film assumption and produce iSCAT-like scattering signatures that are not captured by this framework.

**Operando synchrotron grazing-incidence X-ray diffraction (GIXRD).** Operando GIXRD measurements were performed at the ID31 beamline of the European Synchrotron Radiation Facility (ESRF-EBS, France). A monochromatic X-ray beam with an energy of 75.0 keV and a spot size of 18 × 5 µm² (horizontal × vertical) was used to probe the $Ti_3C_2T_x$ MXene film working electrode under operando conditions. The film was mounted in a PEEK-based three-electrode electrochemical cell, using 3 M $H_2SO_4$ as the electrolyte, carbon cloth as the counter electrode, and an Ag/AgCl reference electrode (Supplementary Fig. S13). The electrochemical cell was mounted on a goniometer to enable precise alignment of the MXene film in grazing incidence conditions. Diffraction patterns were collected using a Pilatus3 CdTe 2M detector positioned at a sample-to-detector distance of 101 cm. Geometric calibration was performed using a $CeO_2$ reference standard. GIXRD measurements were carried out during cyclic voltammetry at a scan rate of 10 mV $s^{-1}$, with data acquisition performed every 10 mV. The resulting data were analyzed by fitting the (*002*) Bragg reflection in reciprocal-space ($Q$–$\chi$) maps, extracting the peak position in $Q$ and $\chi$, as well as the FWHM.

**AFM.** Atomic force microscopy images were acquired in tapping mode under ambient conditions. The resulting height profiles were used to calibrate sthe optical-contrast-to-thickness relation described above (Supplementary Note 3.3).

**Data availability.** The operando iSCAT/IRM image stacks are available at [URL to be added] and the GIXRD data can be found at the ESRF data portal (https://doi.esrf.fr/10.15151/ESRF-ES-1876629818)

**Code availability.** Custom Python scripts for image processing, contrast analysis and derivative computation are available from the corresponding authors upon request.

**Acknowledgements**

We acknowledge the European Synchrotron Radiation Facility (ESRF) for provision of synchrotron radiation facilities and support in using beamline ID31. We acknowledge financial support from the DFG under Germany's Excellence Strategy, EXC 2089/2-390776260, e-conversion Excellence Research Cluster, the Bavarian program Solar Energies Go Hybrid (SolTech) and the Center for Nanoscience (CeNS). M.B. acknowledges financial support from the Marie Skłodowska-Curie Actions PLOBOT (101066396), EXIST Forschungstransfer and the EIC Transition program.

**Author contributions**

F.G., M.B., Y.G., and E.C conceived and designed the project. F.G. and C.G.G. built the iSCAT/IRM setup and F.G. performed the operando optical measurements. F.G. and M. B. designed the iSCAT/IRM experiments and F.G. analyzed the data. R.W. and N.Q.N. synthesized the MXene material and provided ensemble measurements, supervised by Y.G. P.S. and J.D. performed the operando GIXRD measurements at ESRF beamline ID31. P.S., and J.D. analyzed the GIXRD data. F.G. and M.B. wrote the manuscript with input from all authors. M.B. and E.C. supervised and financed the project. All authors discussed the results and commented on the manuscript.

## Competing interests

F.G., C.G.G., M.B., and E.C. have filed a patent (EP4528352 (A1)) for advanced optical microscopy for real-time dynamics of energy materials and C.G.G., M.B. and E.C. are involved in commercializing it through a spin-off project, iNSyT solutions. All other authors declare no competing interests.

## Additional information

Supplementary information is available for this paper.

Correspondence and requests for materials should be addressed to M.B., Y.G. and E.C.